\documentstyle[preprint,aps]{revtex}
\tightenlines

\begin{document}
\newcommand{\beq}{\begin{equation}}
\newcommand{\eeq}{\end{equation}}
\newcommand{\beqa}{\begin{eqnarray}}
\newcommand{\eeqa}{\end{eqnarray}}
\newcommand{\fr}{\frac}
\draft
\preprint{INJE-TP-02-01, hep-th/0201176}

\title{Absorption cross section in de Sitter space}

\author{ Y.S. Myung\footnote{Email-address :
ysmyung@physics.inje.ac.kr}}
\address{Relativity Research Center and School of Computer Aided Science,
Inje University, Gimhae 621-749, Korea}

\maketitle

\begin{abstract}
We study the wave equation for  a minimally coupled massive scalar
 in three-dimensional
de Sitter space. We compute the absorption cross section
 to investigate  its cosmological horizon in the southern diamond.
Although  the absorption cross section is
 not defined exactly,
we can be determined it from the fact that
the low-energy $s(j=0)$-wave absorption cross section for a massless scalar is given
  by the area
of the cosmological horizon. On the other hand, the low-temperature limit of
$j\not=0$-mode absorption cross section is useful
for extracting  information surrounding the cosmological horizon.
 Finally we mention a
computation of the absorption cross section on the CFT-side using the
dS/CFT correspondence.

\end{abstract}
\vfill
Compiled at \today : \number \time.

\newpage

\section{Introduction}
Recently an accelerating universe has proposed to be a way
to interpret the astronomical data of supernova\cite{Per,CDS,Gar}.
The inflation is employed to solve the cosmological flatness and
horizon puzzles arisen in the standard cosmology.
Combining this observation with the need of inflation
 leads to that our universe approaches de Sitter
geometries in both the infinite past and the infinite future\cite{Wit,HKS,FKMP}. Hence it is
very important to study the nature of de Sitter (dS) space and the
dS/CFT correspondence\cite{BOU,STR}.
 However,
there exist  difficulties in studying de Sitter space.
First there is no spatial infinity and   global timelike
Killing vector.  Thus it is not easy to define  conserved  quantities including  mass,
charge and angular momentum appeared in asymptotically  de Sitter space.
Second the dS solution is absent from string theories and thus we
do not   have a definite example  to test the dS/CFT correspondence.
Finally it is hard to define  the $S$-matrix because of the
presence of the cosmological horizon.

We remind the reader that the cosmological horizon is very similar
to the event horizon in the sense that one can define its
thermodynamic quantities using the same way as was done for
the black hole. Two important quantities in the black hole physics
are the Bekenstein-Hawking entropy and the absorption cross
section (=greybody factor). The former relates to the intrinsic
property of the black hole itself, while the latter relates to the
effect of spacetime curvature. Explicitly the greybody factor for
the black hole arises as a consequence of scattering off the
gravitational potential surrounding the horizon\cite{grey1}. For example,
the low-energy $s$-wave greybody factor for a massless scalar has a
universality such that it
is equal to the area of the horizon for all spherically symmetric
 black holes\cite{grey2}. In this work we assume that this universality
 is valid for the cosmological horizon.   The entropy for the cosmological horizon was
 discussed in literature\cite{entropy}. However, as far as we know, there is no
 any explicit computation of the greybody factor for the
 cosmological horizon\footnote{Recently a similar work for four-dimensional
 Schwarzschild-de Sitter black hole appeared in\cite{KOY}. But it considered
 mainly the black hole temperature. Also the absorption rate for the Kerr-de Sitter
 black hole was discussed in\cite{STU}.}.

In this paper we compute the absorption cross section of a massive
scalar in the background of  three-dimensional de Sitter space.
For this purpose we first analyze  the wave equation only in the southern
diamond where the time evolution of  waves is properly defined.
We wish to calculate the outgoing flux near $r=0$. And then we
compute the outgoing flux  by
using the matching  region of overlapping  validity  near the cosmological horizon
of $r_c=1$. Here we follow the conventional approach for a computation of the
greybody factor of the black hole.

The organization of this paper is as follows. In section II we
briefly review the wave equation in de Sitter space. We perform
a potential analysis to study  the asymptotic region by
introducing a tortoise coordinate $r^*$ in section III.
In section IV we calculate the  flux at $r=0,1$ to obtain the greybody
factor.
Finally we discuss our results in section VI.

\section{ wave equation in de Sitter space}

We start with the wave equation for a massive scalar
\beq
(\nabla^2 -m^2) \Phi=0
\label{2eq1}
\eeq
in the background of three-dimensional de Sitter space expressed in the static
coordinates
\beq
ds_{dS_3}^2=-\Big(1- \fr{r^2}{\ell^2} \Big) dt^2 +
\Big(1- \fr{r^2}{\ell^2} \Big)^{-1}dr^2 +r^2
d\phi^2.
\label{2eq2}
\eeq
Here  $\ell$
is the curvature radius of de Sitter space and
hereafter we set $\ell=1$ for simplicity unless otherwise stated.
The above metric is singular at the cosmological horizon, which
divides space into four regions. There are two regions with $0\le r \le1$
which correspond to the causal diamonds of observers at the north
and south poles : northern diamond (ND) and southern diamond (SD).
Two regions with $1<r<\infty$ containing the future-null infinity ${\cal I}^+$ and
past-null infinity  ${\cal I}^-$ are called future triangle (FT)
and past triangle (PT). On ${\cal I}^\pm$ the metric (\ref{2eq2})
is conformal to the cylinder.
A timelike Killing vector $\fr{\partial}{\partial t}$ is
future-directed only in the southern diamond. To obtain the
greybody factor, we have to get a definite wave propagation
as  time evolves. Hence in this work we confine ourselves to the southern diamond.
This means that our working space is compact, in contrast to the
case of the black hole. This will give rise to an ambiguity to
derive an exact form of the greybody factor.

In connection with the dS/CFT correspondence,
one may classify the mass-squared $m^2$ into three cases\cite{STR} :
$m^2\ge1,~~0<m^2<1,~~m^2=0$.
For a massive scalar with  $m^2\ge1$, one has a non-unitary CFT~\footnote{
Although this belongs to one of examples of non-unitary theories
that are dual to well-behaved stable
bulk theory, but the connection between the bulk theory and its non-unitary boundary theory
is not understood clearly  up to now.}. A scalar with mass $0<m^2<1$
can be related to a unitary  CFT.  A massless scalar with $m^2=0$
is special and it would be treated separately.
For our purpose we consider $m^2$ as a parameter at the beginning.
Assuming a mode solution
\beq
\Phi(r,t,\phi)=f(r) e^{-i \omega t} e^{ i j \phi},
\label{2eq3}
\eeq
Eq.(\ref{2eq1}) leads to the differential equation for
$r$\cite{BMS}
\beq
(1-r^2)f''(r) +( \fr{1}
{r} -3r) f'(r) + \Big( \fr{\omega^2}{1-r^2}
-\fr{j^2}{r^2} -m^2 \Big) f(r)=0,
\label{2eq4}
\eeq
where the prime ($'$) denotes the differentiation with respect to
its argument.

\section{potential analysis}

We observe from Eq.(\ref{2eq4}) that
 it is not easy to understand how scalar waves propagate in the southern
diamond. In order to do that, we must transform the wave equation
into the Schr\"odinger-like equation using a tortoise
coordinate $r^*$\cite{ML}.  Then we can get
wave forms in asymptotic regions of $r^*\to \pm \infty$ through a potential analysis.
We introduce  $r^*=g(r)$ with $ g'(r)=1/r(1-r^2)$ to transform  Eq.(\ref{2eq4})
into the Schr\"odinger-like equation with the energy $E=\omega^2$

\beq
-\fr{d^2}{d r^{*2}} f + V(r)f=  E f
\label{3eq1}
\eeq
with a potential
\beq
V(r)= \omega^2 + r^2(1-r^2) \Big[ m^2 + \fr{j^2}{r^2} -\fr{\omega^2}{1-r^2}
\Big].
\label{3eq2}
\eeq
Considering $r^*=g(r)= \int g'(r) dr$, one finds
\beq
r^*= \ln r - \fr{1}{2} \ln(1-r^2),~~ e^{2r^*} =\fr{r^2}{1-r^2},~~
r^2=\fr{e^{2r^*}}{1+e^{2r^*}}.
\label{3eq3}
\eeq
Here we confirm that $r^*$ is a tortoise
coordinate such that $r^* \to -\infty (r \to 0)$, whereas $r^* \to \infty (r \to
1)$. Let us express the potential as a function of $r^*$
\beq
V(r^*) = \omega^2 + \fr{e^{2r^*}}{(1+e^{2r^*})^2} \Big[ m^2 +
\fr{1+ e^{2r^*}}{e^{2 r^*}} j^2 - (1 + e^{2r^*}) \omega^2 \Big].
\label{3eq4}
\eeq
For $m^2=1,j=0,\omega=0.1$, the shape of this takes a  potential barrier ($\frown$) located at
$r^*=0$. On the other hand, for all non-zero $j$,  one finds the
potential step ($\lnot$) with its height $\omega^2+j^2$ on the left-hand side of $r^*=0$.
 All potentials decrease exponentially to zero as $r^*$
increases on the right-hand side.
From the quantum mechanics we always have  a well-defined wave near
the cosmological horizon of $r^*=\infty$.
But near the coordinate origin of $r^*=-\infty~(r=0)$
it is not easy to develop a  genuine traveling wave.

Near $r=0~(r^*=-\infty)$ one finds the equation
\beq
\fr{d^2}{d r^{*2}} f_{(-\infty)} -j^2 f_{(-\infty)}=0.
\label{3eq5}
\eeq
This  gives us a  solution
\beq
f_{(-\infty)}(r^*) =A e^{j r^*} + B e^{-j r^*}
\label{3eq6}
\eeq
which is equivalently rewritten by making use of Eq.(\ref{3eq3}) as
\beq
f_{r=0}(r) =A r^j + \fr{B}{ r^j}.
\label{3eq7}
\eeq
The first term corresponds to a normalizable mode at $r=0~(r^*=-\infty)$, while the
second is a non-normalizable, singular mode. As one discards the second term
in Eq.(\ref{3eq7})
for  calculating the Bogoliubov transformation~\cite{BMS}, the first term
is needed  for our purpose. Hence we set $B=0$.
Also we observe  that  angular momentum modes play the important
role in the computation of the flux.

On the other hand,
near the cosmological horizon $r_c=1(r^*=\infty)$ one obtains a differential equation
\beq
\fr{d^2}{d r^{*2}} f_{\infty} +\omega^2 f_{\infty}=0.
\label{3eq8}
\eeq
This  has  a  solution
\beq
f_{\infty}(r^*) =C e^{-i \omega r^*} + D e^{i \omega r^*}
\label{3eq9}
\eeq
which is equivalently rewritten as
\beq
f_{r=1}(r) =C(1- r^2)^{\fr{i\omega}{2}} + D(1- r^2)^{-\fr{i\omega}{2}} .
\label{3eq10}
\eeq
The first wave (second wave) in Eq.(\ref{3eq9}) together with $e^{-i \omega t}$
implies  the ingoing (outgoing) waves across the cosmological
horizon.
In order to obtain a connection between $A$ and $C,~D$, let us
consider the case of $j^2 > m^2$   with $V\approx  V_0= \omega^2+j^2$
for $ -\infty < r^* \le 0$ and $V \approx 0$ for $ 0\le r^* < \infty$ with the energy
$E=\omega^2 <V_0$. Also it corresponds to the problem for a
 potential step $V_0$ with $0<E<V_0$ in the quantum mechanics. Requiring
the conservation of flux at $r^*=0$ leads to the asymptotic relations
: $C \approx (j-i \omega)A/(-2i \omega),~D \approx (j+i \omega)A/(2i \omega)$.
Although the flux on the left hand side of $r^*=0$ is zero due to
the real function of Eq.(\ref{3eq6}), the flux of the right hand side
is not zero because of the traveling wave nature of Eq.(\ref{3eq9}).

Up to now we obtain  asymptotic forms of a scalar wave which
propagates in the southern diamond of de Sitter space.
In order to calculate the absorption cross section, we need to know
an explicit form of wave  propagation in $0 \le r \le 1$. This can be  achieved
only when
solving the differential equation (\ref{2eq4}) explicitly.

\section{Flux calculation}

In order to solve equation (\ref{2eq4}), we first transform it into
a hypergeometric equation using $z=r^2$. Here  the working space still remains
unchanged as $0 \le z \le 1$ covering  the southern diamond.
This equation  takes a form

\beq
z(1-z)f''(z) +(1-2z) f'(z) + \fr{1}{4} \Big( \fr{\omega^2}{1-z}
-\fr{j^2}{z} -m^2 \Big) f(z)=0.
\label{4eq1}
\eeq
Here one finds two poles at $z=0,1 (r=0,1)$ and so  makes a further
transformation  to cancel these by choosing
\beq
f(z)=z^\alpha (1-z)^\beta w(z),~~
\alpha_{\pm}=\pm \fr{j}{2},~~\beta_{\pm}=\pm i\fr{\omega}{2}.
\label{4eq2}
\eeq
First we choose the case of $\alpha_+$ and $~\beta_+$.
Then we obtain a hypergeometric equation
\beq
z(1-z)w''(z) + [c-(a+b+1)z] w'(z) -a b~ w(z) =0
\label{4eq3}
\eeq
where $a,b$ and $c$ are given by
\beq
 a= \fr{1}{2} ( j + i \omega +h_+),~~ b= \fr{1}{2} ( j + i \omega
 +h_-),~~ c= 1 +j
 \label{4eq4}
 \eeq
 with
 \beq
 h_+= 1+ \sqrt{1-m^2} \Big(=1+ i \sqrt{m^2-1} \Big),~~ h_-= 1- \sqrt{1-m^2} \Big(=1-i
 \sqrt{m^2-1} \Big).
 \label{4eq5}
 \eeq
 Considering $j=$ integer,
 one regular solution near $z=0$ to Eq.(\ref{4eq1}) is given
 by\cite{AS}
 \beq
 f_+(z)
 =A z^{j/2}(1-z)^{i\omega/2} F(a,b,c;z)
 \label{4eq6}
 \eeq
with an unknown constant $A$. Also there is the other solution with a
logarithmic singularity at $z=0$ as $\tilde f_+(z)=
 \tilde A z^{j/2}(1-z)^{i \omega/2}[F(a,b,c;z) \ln z +\cdots]$.
However, both solutions have vanishing
flux at $z=0$, because the relevant part ($z^{j/2}$)
is not complex but real.

Concerning  the case of $\alpha_-$ and $~\beta_+~(c=1-j)$, one has one regular solution
\beq
\label{4eq7}
f_-(z)=Ez^{j/2}(1-z)^{i\omega/2} F(a+j,b+j,1+j;z)
\eeq
and the other
singular solution is given by
 $\tilde f_-(z)=\tilde E z^{j/2}(1-z)^{i\omega/2} [F(a+j,b+j,1+j;z) \ln z
+\cdots]$. These have  essentially the same behaviour at $z=0$ as in the
case of $\alpha_+~(c=1+j)$.

Now we are in a position to calculate an  outgoing flux at $z=0$ which is defined   as
\beq
{\cal F}(z=0)= 2\fr{2 \pi}{i} [f^*z\partial_z f-f z\partial_z
f^*]|_{z=0}.
\label{4eq8}
\eeq
For any kind of real functions near $z=0$ including $f_{\pm}$ and $\tilde f_{\pm}$,
 the outgoing/ingoing flux is given by
\beq
{\cal F}_{out/in}(z=0)=0.
\label{4eq9}
\eeq
Hence we choose a regular  solution of Eq.(\ref{4eq6}) for
further calculation.
To obtain a flux at the horizon of $z=1(r=1)$, we first have to use a formula : $F(a,b,c;z)=
\fr{\Gamma(c)\Gamma(c-a-b)}{\Gamma(c-a)\Gamma(c-b)} F(a,b,a+b-c+1;1-z)
+ \fr{\Gamma(c)\Gamma(a+b-c)}{\Gamma(a)\Gamma(b)}(1-z)^{c-a-b} F(c-a,c-b,-a-b+c+1;1-z).$
Using $1-z \sim e^{-2r^*}$ near $z=1$, one finds from Eq.(\ref{4eq6}) the
following form:

\beq
f_{0\to1}\equiv f_{in} + f_{out}=
 H_{\omega,j} e^{-i \omega r^*} +H_{-\omega,j} e^{i \omega
r^*}
\label{4eq10}
\eeq
where
\beq
H_{-\omega,j}= A\alpha_{-\omega,j},~~
\alpha_{-\omega,j}= \fr{\Gamma(1+j)\Gamma(i\omega) 2^{i \omega}}
{\Gamma[(j +i \omega +h_+)/2)] \Gamma[(j +i \omega +h_-)/2)]}.
\label{4eq11}
\eeq
Then we match Eq.(\ref{3eq9}) with Eq.(\ref{4eq10}) to yield $C=H_{\omega,j}$
and $D=H_{-\omega,j}$ near the cosmological horizon.
Finally we calculate its outgoing flux at $z=1(r^*=\infty)$ as
\beq
{\cal F}_{out}(z=1)= \fr{2 \pi}{i} [f_{out}^*\partial_{r^*} f_{out}-f_{out}
\partial_{r^*}
f_{out}^*]|_{r^*=\infty} = 4 \pi \omega A^2
|\alpha_{-\omega,j}|^2.
\label{4eq12}
\eeq
If one uses Eq.(\ref{4eq7}) instead of Eq.(\ref{4eq6}),
its outgoing  wave
near $z=1$ is given by $B\alpha_{-\omega,j}e^{i\omega r^*}$.
This is the same flux as in Eq.(\ref{4eq12}) except of
replacing $A$ by $B$.

\section {absorption cross section}

An absorption coefficient by the cosmological horizon is defined formally by
\beq
{\cal A} = \fr{ {\cal F}_{out}(z=1)}{{\cal F}_{out}(z=0)}.
\label{5eq0}
\eeq
However, we have zero-normalization of ${\cal F}_{out}(z=0)=0$ for
de Sitter wave propagation. In this case one cannot define the
absorption cross section. Instead  we propose an unknown normalization
${\cal F}_{out}(z=0) =F$, where $F$ will be determined by referring
the absorption cross section for  the low-energy $s(j=0)$-wave.
Up to now we do not
insert the curvature radius $\ell$ of dS$_3$ space.
The correct absorption coefficient can be recovered when replacing $\omega(m)$ with
$ \omega\ell(m\ell)$.
Then the absorption cross section in three dimensions is defined
by
\beq
\sigma_{abs}= \fr{{\cal A}}{\omega} =
\Big[\fr{4\pi \ell A^2}{F}\Big] |\alpha_{-\omega\ell,j}|^2
\label{5eq1}
\eeq
where
\beq
|\alpha_{-\omega\ell,j}|^2=\fr{|\Gamma(1+j)|^2|\Gamma(i\omega\ell)|^2 }
{|\Gamma[(j + i\omega\ell +h_+)/2)]|^2
|\Gamma[(j + i\omega\ell +h_-)/2]|^2}.
\label{5eq2}
\eeq
We calculate the low-energy $ s(j=0)$-wave cross section for a massless scalar
 to determine $F$\cite{grey2}.
This is given by
\beq
\sigma_{abs,j=0}^{m\ell=0,\omega \ell \ll 1}= \pi \ell \fr{A^2}{F}
\to {\cal A}_{ch}
\label{5eq3}
\eeq
with the area of the cosmological horizon ${\cal A}_{ch}=2 \pi
\ell$. Hence the unknown flux ${\cal F}(z=0)$  is determined as
$F=A^2/2$. We wish to classify the absorption cross section
according to values of $m\ell$.
\subsection{$m\ell=0$ case}

This case corresponds to the absorption cross section of a
massless scalar. Using $h_+=2,h_-=0$, one finds
\beqa
\sigma_{abs}^{m\ell=0}&=&\Big[8\pi \ell\Big]
\fr{(j!)^2\pi}{ \omega\ell \sinh[\pi\omega\ell]} \nonumber \\
 && \times \fr{1}{|\Gamma(1+j/2 + i\omega\ell/2)|^2|\Gamma(j/2 + i\omega\ell/2)|^2}.
\label{5eq4}
\eeqa
For $j=1$, this takes the form
\beqa
\sigma_{abs,j=1}^{m\ell=0}&=&
\fr{16 {\cal A}_{ch}}{ \pi \omega\ell \sinh[\pi\omega\ell]} \nonumber \\
 && \times
 \fr{\Big(\cosh[\pi\omega\ell/2]\Big)^2}{(1+(\omega\ell)^2)}.
\label{5eq5}
\eeqa
In order to get a definite expression for the absorption cross section,
let us consider the low-energy scattering with small $E=(\omega\ell)^2<<1$
(for example, $\omega\ell=0.1$). Introducing the temperature of
the cosmological horizon $T_{ch}=\fr{1}{2\pi\ell}$, this limit implies
$\omega<<2T_{ch}$. On the other hand there exist the low-temperature
limit of $\omega\ell>>1(\omega>>2T_{ch})$~\footnote{ In the study of the cosmological
horizon, this high frequency approximation is very useful for obtaining the absorption
cross section\cite{ST}. Considering $\omega= 2\pi \nu \sim 1/\lambda$, this approximation
implies $\lambda << \ell$, which means that the size of the cosmological horizon is very bigger
than the wavelength of a test scalar.
Hence  the scattering of a scalar off the cosmological horizon
is well-defined in the low-temperature limit (that is, high frequency approximation).}.
In the low-energy limit of $\omega\ell << 1$, $j=1$-mode reduces to
\beq
\sigma_{abs,j=1}^{m\ell=0,\omega\ell<<1}=
\fr{16}{ \pi^2} \fr{{\cal A}_{ch}}{(\omega\ell)^2}.
\label{5eq6}
\eeq
On the other hand, its low-temperature limit  is given by
\beq
\sigma_{abs,j=1}^{m\ell=0,\omega\ell>>1}=
\fr{8}{ \pi} \fr{{\cal A}_{ch}}{(\omega\ell)^3}.
\label{5eq6'}
\eeq
For $j=2$, this leads to
\beqa
\sigma_{abs,j=2}^{m\ell=0}&=&
\fr{64{\cal A}_{ch} \pi}{\omega\ell \sinh[\pi\omega\ell]} \nonumber \\
 && \times \fr{\Big(\sinh[\pi\omega\ell/2]\Big)^2}{(4+(\omega\ell)^2)(\pi\omega\ell/2)^2}.
\label{5eq7}
\eeqa
In the low-energy limit of $\omega\ell << 1$, this reduces to
\beq
\sigma_{abs,j=2}^{m\ell=0,\omega\ell<<1}=
 \fr{16{\cal A}_{ch}}{(\omega\ell)^2}.
\label{5eq8}
\eeq
But its low-temperature limit is
\beq
\sigma_{abs,j=2}^{m\ell=0,\omega\ell>> 1}=
\fr{128}{ \pi} \fr{{\cal A}_{ch}}{(\omega\ell)^5}.
\label{5eq8'}
\eeq

\subsection{$0<m\ell<1$}
In this case there is no further expression for the absorption cross section
 because of the
real values of $h_\pm=1\pm \sqrt{1-(m\ell)^2}$.

\subsection{$m\ell=1$ case}
Here we have the absorption cross section
\beqa
\sigma_{abs}^{m\ell=1}&=&\
\Big[8 \pi \ell \Big]\fr{(j!)^2 \pi}{ \omega\ell \sinh[\pi\omega\ell]} \nonumber \\
 && \times \fr{1}{\Big(|\Gamma[(j+1+
 i\omega\ell)/2]|^2\Big)^2}.
\label{5eq9}
\eeqa
In the low-energy limit of $\omega\ell << 1$, the cross section for $j=1$ reduces to
\beq
\sigma_{abs,j=1}^{m\ell=1,\omega\ell <<1}=
\fr{4{\cal A}_{ch}}{(\omega\ell)^2},
\label{5eq10}
\eeq
while its low-temperature limit is
\beq
\sigma_{abs,j=1}^{m\ell=1,\omega\ell>>1}=
\fr{2}{ \pi} \fr{{\cal A}_{ch}}{(\omega\ell)^3}.
\label{5eq10'}
\eeq
The low-energy  cross section for $j=2$ is given by
\beq
\sigma_{abs,j=2}^{m\ell=1,\omega\ell << 1}=
\fr{256}{ \pi^2} \fr{{\cal A}_{ch}}{(\omega\ell)^2}.
\label{5eq11}
\eeq
But the low-temperature cross section leads to
\beq
\sigma_{abs,j=2}^{m\ell=1,\omega\ell>> 1}=
\fr{128}{ \pi} \fr{{\cal A}_{ch}}{(\omega\ell)^5}.
\label{5eq11'}
\eeq

\subsection{$m\ell >1$}

In this case we have to use another expression of $h_\pm=1\pm i\sqrt{(m\ell)^2-1}$.
For even $j$, one finds the absorption cross section
\beqa
\sigma_{abs,j=even}^{m\ell > 1}&=&\Big[ 8 \pi \ell \Big]
\fr{(j!)^2\cosh[\pi y_1/2]\cosh[\pi y_2/2]}{\pi \omega\ell \sinh[\pi\omega\ell]} \nonumber \\
 && \times \fr{1}{|(j-1)/2+ i y_1/2|^2 \cdots |1/2+ i y_1/2|^2
|(j-1)/2+ i y_2/2|^2 \cdots |1/2+ i y_2/2|^2}
\label{5eq12}
\eeqa
with $y_1=\omega \ell +\sqrt{(m\ell)^2-1} $
and $y_2=\omega\ell -\sqrt{(m\ell)^2-1}$.
For odd $j$, one has
\beqa
\sigma_{abs,j=odd}^{m\ell > 1}&=& \Big[8 \pi \ell\Big]
\fr{(j!)^2\sinh[\pi y_1/2] \sinh[\pi y_2/2]}
{\pi \omega\ell (y_1/2)(y_2/2)\sinh[\pi\omega\ell]} \nonumber \\
&& \times \fr{1}{|(j-1)/2+ i y_1/2|^2 \cdots |1+ i y_1/2|^2
|(j-1)/2+ i y_2/2|^2 \cdots |1+ i y_2/2|^2}.
\label{5eq13}
\eeqa
\section{discussion}

We calculate the absorption cross section of a minimally coupled
scalar which propagates in the southern diamond of
three-dimensional de Sitter space. We expect that
higher-dimensional cases have similar behaviors because their wave
equation takes the nearly same form as in the three-dimensional
case in Eq.(\ref{2eq4}).

One of the striking results is that the low-energy $s(j=0)$-wave
absorption of a massless scalar is not defined exactly within our approach. This
mainly depends on being  unable to calculate its non-zero  flux at
$r=0(z=0)$. However, using  a massless scalar propagation in the symmetric black holes
whose cross sections are  the area of the event
horizon of black hole \cite{grey2}, we  calculate the low-energy $s(j=0)$-wave
absorption  Eq.(\ref{5eq3}) of a massless scalar in the background of de Sitter horizon.

On the other hand,  $j\not=0$-modes of a scalar wave
play the important role in extracting   information about the
cosmological horizon compared to the black holes.
We expect  from the black hole analysis that
the low-energy limit ($\omega R \to 0 $) of the $l\not=0$-angular mode absorption cross section
is proportional roughly to $(\omega R)^{4l} {\cal A}_{7BH}$ for the  7D black holes which
was derived from D3-branes\cite{GH}. For 5D black holes, it is
proportional to $(\omega r_o)^{2l}{\cal A}_{5BH}$\cite{MS}.
 However, one finds from Eqs.(\ref{5eq6}), (\ref{5eq8}),
 (\ref{5eq10}) and (\ref{5eq11}) that
those for $j\not=0$ in the low-energy limit of $\omega\ell<<1$ are given by $1/(\omega\ell)^2$
which implies that
the absorption cross section is greater than the area of the cosmological horizon.
This is not the promising case that we want to get.
From the potential analysis
it conjectures that for $j\not=0$,  the low-energy absorption cross section
with $E=(\omega\ell)^2<<1$
is negligible because a potential step with its height
$V_0=j^2+(\omega\ell)^2>>E$
is present on the left hand side.
Thus it seems that taking a low-energy limit of the de Sitter absorption
cross section for $j\not=0$-mode is meaningless
because it is always greater than ${\cal A}_{ch}$. In  order  to obtain a
finite absorption cross section,
$\omega\ell$ should be large such as $\omega\ell >> 1$. Actually this corresponds to
  the low-temperature limit of $\omega>>2T_{ch}$.
The low-temperature limit (that is, high frequency approximation)
 is  meaningful in the study of de Sitter space
 since its cross section appears less than ${\cal A}_{ch}$.
We find from Eqs.(\ref{5eq6'}), (\ref{5eq8'}), (\ref{5eq10'}) and (\ref{5eq11'})
that the absorption cross section decreases as $j$ increases.
This is consistent with the potential analysis which states that for given energy
 $E=\omega^2$, the potential increases as $j$ increases.
As a result, the low-temperature limit  of
$j\not=0$-mode absorption cross section will be used to extract
information about the cosmological horizon.

At this stage we wish to comment on   the other approach of
Suzuki and Takasugi in deriving the
absorption cross section in de Sitter space\cite{ST}.
Actually a definition of  absorption probability
Eq.(16) in ref.\cite{ST} can be applicable for the non-zero
spin field. Hence their approach may be  problematic to a spin-zero scalar.
However, an interesting fact that
the absorption cross section of the cosmological horizon  can be well-defined
in the low-temperature limit  is
checked by two approaches independently.

Also we wish to mention the difference between the Bogoliubov
coefficient and the greybody factor. In calculating the Bogoliubov
coefficient, we have to define two different
vacuums by introducing global~\cite{SS} or Kruskal coordinates~\cite{BMS}.
In calculating the Bogoliubov
coefficients, one needs only a normalizable mode in
Eq.(\ref{4eq6}). Instead the entire space (SD,ND,FT,PT) normalizable solutions are used
for finding thermal state of the cosmological horizon by matching
them across the the cosmological horizon.
 Then we find
the thermal nature of the cosmological horizon clearly, for example,
 Planck (black body) spectrum with temperature.  In this
case the fine effect of scattering is usually neglected
~\cite{BD}.

However, de Sitter space gives us a potential
surrounding the cosmological horizon and it gives rise to a
scattering for a test scalar wave which propagates from $r=0$ to
$r=\ell$.
We need  each non-zero flux at $r=0,1$ in the southern diamond
to calculate the transmission (absorption) coefficient. But
we have  zero-flux at $r=0$. Since the
working space is compact and it includes the origin of coordinate
$(r=0$), it may be problematic to derive an exact form of the absorption cross
section in this way.
In this work, calculating the absorption cross section , we use
only
a normalizable mode Eq.(\ref{4eq6}) within the southern diamond (SD).
Hence  we do not have
enough information to determine the absorption cross section in the bulk spacetime approach.
Fortunately,
using the universality for s-wave cross section in the black hole
background, we could determine the absorption cross section.
This also happens in the study of the AdS-black holes\cite{flux}.

Finally we wish to mention that the bulk absorption cross section can be
also calculated from the two-point function of CFT defined on the boundary if one assumes the
AdS/CFT correspondence to realize the holographic principle\cite{TEO}.
Hence we hope that our results for bulk dS space can be recovered from
the CFT approach \cite{STR,SV,AWS} by
making use of  the dS/CFT correspondence to realize the assumed dS holography.
Also the dS complementarity principle may help to remedy  the
undetermined normalization factor in Eq.(\ref{5eq2})\cite{DDO}.
Consequently the  exact form of the  absorption cross
section for the cosmological horizon  will be found if one gets the
CFT results.

\section*{Acknowledgement}
We thank   Hyung Won Lee for helpful discussions.
This work was supported in part by  KOSEF, Project
Nos. R01-2000-000-00021-0 and R02-2002-000-00028-0.


\begin{references}

\bibitem{Per} S. Perlmutter et al.(Supernova Cosmology Project), Astrophys. J.
{\bf 483}, 565(1997)[astro-ph/9608192].

\bibitem{CDS}R. R. Caldwell, R. Dave, and  P. J. Steinhard, Phys. Rev. Lett.
{\bf 80}, 1582(1998)[astro-ph/9708069].

\bibitem{Gar}P. M. Garnavich et al.,  Astrophys. J. {\bf 509}, 74(1998)[astro-ph/9806396].

\bibitem{Wit} E.Witten, ``Quantum Gravity in de Sitter Space",
hep-th/0106109.

\bibitem{HKS}S. Hellerman, N. Kaloper, and L. Susskind, JHEP {\bf 0106}, 003(2001)[hep-th/0104180].

\bibitem{FKMP}  W. Fischler, A. Kashani-Poor,
R. McNees, and  S. Paban, JHEP {\bf 0107}, 003 (2001)[hep-th/0104181].

\bibitem{BOU} R.Bousso, JHEP {\bf 0011}, 038 (2000)[hep-th/0010252];
R. Bousso, JHEP {\bf 0104}, 035 (2001)[hep-th/0012052];
S. Nojiri and D. Odintsov, Phys.Lett. {\bf B519}, 145 (2001)[
hep-th/0106191]; D. Klemm, ``Some Aspects of the de Sitter/CFT
correspondence", hep-th/0106247 ;
M. Spradlin, A. Strominger, and A. Volovich, ``Les
Houches Lectures on De Sitter Space", hep-th/0110007;
S. Cacciatori and D. Klemm, ``The Asymptotic
Dynamics of de Sitter Gravity in three Dimensions", hep-th/0110031;
 A. C. Petkou and G. Siopsis, ``dS/CFT correspondence
on a brane", hep-th/0111085.
V. Balasubramanian, J. de Boer, and D. Minic,
``Mass, Entropy and Holography in Asymptotically de Sitter
Spaces", hep-th/0110108;  R. G. Cai, Y. S. Myung, and Y. Z. Zhang, ``Check of
the Mass Bound Conjecture in de Sitter Space", hep-th/0110234;
 Y. S. Myung, Mod. Phys. Lett.{\bf A16}, 2353(2001)[hep-th/0110123];
 R. G. Cai, ``Cardy-Verlinde Formula and
Asymptotically de Sitter Spaces", hep-th/0111093;
 A. M. Ghezelbach and R. B. Mann, JHEP {\bf 0201}, 005 (2002)[hep-th/0111217];
 M. Cvetic, S. Nojiri, and S.D. Odintsov,
``Black Hole Thermodynamics and Negative Entropy in deSitter and
Anti-deSitter Einstein-Gauss-Bonnet gravity", hep-th/0112045;
Y. S. Myung, ``Dynamic dS/CFT correspondence using the brane
cosmology", hep-th/0112140; R. G. Cai, ``Cardy-Verlinde Formula and
Thermodynamics of Black Holes in de Sitter Spaces", hep-th/0112253.

\bibitem{STR} A. Strominger, JHEP {\bf 0110}, 034 (2001)[hep-th/0106113];
 A. J. Tolley and N. Turok, ``Quantization of the
massless minimally coupled scalar field and the dS/CFT
correspondence", hep-th/0108119.

\bibitem{grey1} C. Callan, S. Gubser, I. Klebanov, and A.
Tseytlin, Nucl. Phys. {\bf B489}, 65(1997)[hep-th/9610172]; M. Karsnitz and I.
Klebanov, Phys. Rev. {\bf D56}, 2173(1997)[hep-th/9703216]; B. Kol and A.
Rajaraman, Phys. Rev. {\bf D56}, 983(1997)[hep-th/9608126];
M. Cvetic and F. Larsen, Nucl. Phys. {\bf B506},
107(1997)[hep-th/9706071].


\bibitem{grey2} A. Dhar, G. Mandal, and S. Wadia
,Phys. Lett. {\bf B388}, 51(1996) [hep-th/9605234];
 S. Das, G. Gibbons and S. Mathur, Phys. Rev. Lett. {\bf 78}, 417(1977)[hep-th/9609052].



\bibitem{entropy}  M. I. Park, Phys.Lett. {\bf B440}, 275(1998)
 [hep-th/9806119]; M. Banados, T. Brotz and M. Ortiz,  Phys. Rev. {\bf D59},
 046002(1999)[hep-th/9807216]; W. T. Kim,
 Phys. Rev. {\bf D59}, 047503(1999)[hep-th/9810169]; F. Lin and Y. Wu,
 Phys. Lett. {\bf B453}, 222 (1999)[hep-th/9901147].

\bibitem{KOY} W. T. Kim, J.J. Oh, and K. H.
Yee, ``Scattering amplitudes and thermal temperatures of the
Schwarzschild-de Sitter black holes", hep-th/ 0201117.


\bibitem{STU} H. Suzuki, E. Takasugi, and H. Umetsu, Prog. Theor.
Phys. {\bf 103}, 103 (2000)[gr-qc/9911079].



\bibitem{ML} H. W. Lee and Y. S. Myung, Phys. Rev. {\bf D61},
024031(2000)[hep-th/9903054];
H. W. Lee, Y. S. Myung, and J. Y. Kim, Phys. Rev. {\bf D58},
104006(1998)[hep-th/9708099].

\bibitem{BMS} R. Bousso, A. Maloney, and A. Strominger,
``Conformal vacua and entropy in de Sitter space", hep-th/0112218.

\bibitem{AS} M. Abramowitz and I. Stegun, ``Handbook of
Mathematical Functions" (Dover Publicaton ,New York, 1970), p.564.


\bibitem{ST} H. Suzuki and E. Takasugi, Mod. Phys. Lett. {\bf
A11}, 431(1996)[gr-qc/9508068].

\bibitem{GH} S. Gubser and Hashimoto, Commun.Math. {\bf 203},
325(199)[hep-th/9805140].

\bibitem{MS} J. Madacena and A. Strominger, Phys. Rev. {\bf D56},
4975(19970[hep-th/9702015]; S. D. Mathur, Nucl. Phys. {\bf B514},
204(1998)[hep-th/9704156].

\bibitem{SS} H.-T. Sato and H. Suzuki, Mod. Phys. Lett. {\bf A9},
3673 (1994)[hep-th/9410092].

\bibitem{BD} N. D. Birrell and P. C. W. Davies, {\it Quantum
fields in curved space} (Cambridge Univ. Press, New York, 1982) p.
260.
\bibitem{flux} D. B. Birmingham, I. Sachs and S. Sen, Phys. Lett.
{\bf B413}, 281(1997); H.W. Lee, N. J. Kim, and Y. S. Myung,
Phys. Rev. {\bf D58}, 084022 (1988)[hep-th/9803080];
H.W. Lee, N. J. Kim, and Y. S. Myung,
Phys. Lett. {\bf B441}, 83(1988)[hep-th/9803227];
H.W. Lee and Y. S. Myung,
Phys. Rev. {\bf D58}, 104013 (1988)[hep-th/9804095].

\bibitem{TEO} E. Teo, Phys. Lett. {\bf B436},
269(1988)[hep-th/9805014];
Y. S. Myung and H. W. Lee, JHEP {\bf 9910},
009(1999)[hep-th/9904056].

\bibitem{SV} M. Spradlin and A. Volovich, ``Vacuum states and the
S-matrix in dS/CFT", hep-th/0112223.

\bibitem{AWS} E. Abdalla, B. Wang, A. Lima-Santos and W. G. Qiu,
``Support of dS/CFT correspondence from perturbations of three
dimensional spacetime", hep-th/0204030.

\bibitem{DDO} U. H. Danielsson, D. Domert and M. Olsson,
``Miracles and complementarity in de Sitter space",
hep-th/0210198.
















\end{references}
\end{document}